\documentclass[fleqn,usenatbib]{mnras}

\usepackage{newtxtext,newtxmath}

\usepackage[T1]{fontenc}

\DeclareRobustCommand{\VAN}[3]{#2}
\let\VANthebibliography\thebibliography
\def\thebibliography{\DeclareRobustCommand{\VAN}[3]{##3}\VANthebibliography}

\usepackage{graphicx}	
\usepackage{amsmath}	

\def\gs{\mathrel{\hbox{\rlap{\hbox{\lower4pt\hbox{$\sim$}}}\hbox{$>$}}}}
\def\ls{\mathrel{\hbox{\rlap{\hbox{\lower4pt\hbox{$\sim$}}}\hbox{$<$}}}}

\def\mrk79{{Mrk~79}}
\def\iras13{{IRAS~13224--3809}}
\def\1h07{{1H~0707--495}}
\def\izw1{{I~Zw~1}}

\def\ga{{\rm\thinspace gauss}}

\title[Probing the Eclipse of NGC~6814]{A Colourful Analysis: Probing the Eclipse of the Black Hole and Central Engine in NGC 6814 Using X-ray Colour-Colour Grids}

\author[B. Pottie et al.]{
B. Pottie,$^{1}$\thanks{E-mail: Benjamin.Pottie@smu.ca},
L. C. Gallo,$^{1}$
A. G. Gonzalez,$^{1}$ and
J. M. Miller$^{2}$
\\
$^{1}$Department of Astronomy and Physics, Saint Mary’s University, 923 Robie Street, Halifax, NS, B3H 3C3, Canada \\
$^{2}$Department of Astronomy, University of Michigan, 500 Church Street Ann Arbor, MI 48109, USA
}

\date{Accepted XXX. Received YYY; in original form ZZZ}

\pubyear{2015}

\begin{document}
\label{firstpage}
\pagerange{\pageref{firstpage}--\pageref{lastpage}}
\maketitle

\begin{abstract}
 Eclipsing of the X-ray emitting region in active galactic nuclei (AGN) is a potentially powerful probe to examine the AGN environment and absorber properties.  Here we study the eclipse data from the 2016 XMM-Newton observation of NGC 6814 using a colour-colour analysis.  Colours (i.e. hardness ratios) can provide the advantage of better time-resolution over spectral analysis alone.  Colour-colour grids are constructed to examine the effects of different parameters on the observed spectral variability during the eclipse.  Consistent with previous spectral analysis, the variations are dominated by changes in the column density and covering fraction of the absorber.  However, during maximum eclipse the behaviour of the absorber changes.  Just after ingress, the eclipse is described by changes in column density and covering fraction, but prior to egress, the variations are dominated by changes in column density alone.  Simulations are carried out to consider possible absorber geometries that might produce this behaviour.  The behaviour is inconsistent with a single, homogeneous cloud, but simulations suggest that multiple clouds, perhaps embedded in a highly ionised halo, could reproduce the results.  In addition, we determine the orbital covering factor (fraction of orbital path-length) based on evidence of several eclipses in the 2016, 64-day \textit{Swift} light curve.  We estimate that $\sim 2-4$~per cent of the orbit is covered by obscuring clouds and that the distribution of clouds is not isotropic. \\
\end{abstract}

\begin{keywords}
galaxies: active -- galaxies: individual: NGC 6814 -- X-rays: galaxies
\end{keywords}



\section{Introduction}

The emission from Active Galactic Nuclei (AGN) is dominated by processes that occur in the regions closest to the central supermassive black hole.  AGN have the capability to emit over the entire electromagnetic spectrum, but X-rays originate in the inner-most regions and their emission is highly variable.  The processes creating X-rays largely occur from interactions between the hot central corona and the inner accretion disc.  There are constituents within the AGN systems, e.g. the warm absorbers, obscurers, and the dusty torus, that give rise to absorption and modify the X-ray emission we see.  Studying the absorption can yield valuable information about the AGN environment and the region where the absorption originates from (\citealt{Gallo2004}; \citealt{Chainakun2017}; \citealt{Alston2020}; \citealt{Miller2010}).   In the event of eclipses, we can even estimate sizes of the X-ray source, which is important since this region is not resolvable with current detectors (\citealt{Risaliti2007}; \citealt{Risaliti2009b}; \citealt{Risaliti2011}; \citealt{Turner2018}; \citealt{Zoghbi2019}).

Significant X-ray variability in AGN can be attributed to the absorbers within the system, rather than intrinsic to the continuum source.  Many AGN like NGC 1365, 3327, and 3783, have shown eclipses by some obscurer(s) (\citealt{Risaliti2007}; \citealt{Risaliti2009a}; \citealt{Walton2013}; \citealt{Brenneman2013}; \citealt{Turner2018}; \citealt{George1998}; \citealt{DeMarco2020}; \citealt{Constanzo2022}).  Often times, such an obscurer partially covers the source (\citealt{Holt1980}; \citealt{Tanaka2004}; \citealt{Gallo2015}; \citealt{Kara2021}), with the fraction denoted by the covering fraction parameter.  Variability in these obscurers generally involves changes in the column density and/or the covering fraction of the obscurer as it transits the line-of-sight to the X-ray source, but variations in the ionising state can also occur.  Brightness fluctuations because of the obscurer can be particularly useful at understanding the geometry of such clouds.

Eclipsing events are most evident when comparing the light curve and hardness ratio (HR) curve, simultaneously.  Dips in light curve coincident with hardening in the HR curve generally signify an obscuration event (\citealt{Brenneman2013}).  This happens because the absorption has a larger effect at softer energies.  The variability can occur on all timescales from hours (e.g. NGC~6814, \citealt{Gallo2021}; NGC~4395, \citealt{Nardini2011}) up to years (e.g. NGC~5548, \citealt{Kaastra2014}) giving information about the location of the absorption (e.g. \citealt{Bianchi2009}). 

Hardness ratios (colours) are regularly employed in X-ray astronomy to study spectral shapes.  Colours are particularly useful in situations where photon detections are very limited and spectral analysis is challenging.  They were widely applied to low-count data in the ROSAT All Sky Survey (RASS, \citealt{Schartel1996}), to classify objects (e.g. as absorbed or unabsorbed) and to measure photon indices.

The motivation to use colour-colour grids to study in this work comes from \cite{Grinberg2020}.  They used colour-colour grids to distinguish different models of high mass X-ray binaries (HMXBs).  Applying these grids to Cyg X-1, they could understand the influence of different parameters and break the degeneracy between different models.  The colour analysis could be used to discern the effects of changing covering fraction, column density, photon index, and different ionization types (neutral, warm, etc.) in partial coverers.

Similarly, colour-colour grids were used in the work of \cite{Carpano2005} and \cite{Nowak2011}.  The former applied such grids to point sources in the optical disk of NGC 300, allowing them to determine the model that described all but one of the point sources.  The latter applied grids to the dust-scattering halo of Cyg X-1, and found success in discerning some properties of the X-ray source.  Building on the success of these mentioned works and others, here we adopt colour-colour grids to study the eclipse in NGC~6814. 

NGC 6814, is a Seyfert 1.5  AGN ($z=0.00521$) with a black hole mass of M$_{BH}$ = 10$^{7.038^{+0.056}_{-0.058}}$~M$_{\odot}$ (\citealt{Bentz2015}).  It possesses rapid X-ray variability on all timescales and it might exhibit occasional eclipses (e.g. \citealt{Gallo2021}; \citealt{Leighly1994}).  The 2016 eclipse, was captured by \textit{XMM-Newton} in its entirety (\citealt{Gallo2021}).  In a continuous observation, the ingress, low-state, and egress were observed without interruption.  Capturing an entire eclipse in an observation is rather unique, and it allowed for the determination of the size, shape and the location of the obscurer causing the eclipse (\citealt{Gallo2021}).  The eclipse was observed to happen on day-long timescales, something that is consistent with an obscurer located sufficiently close to the black hole (e.g. the broadline region, BLR) (\citealt{Elvis2004}; \citealt{Risaliti2009b}; \citealt{Puccetti2007}; \citealt{Bianchi2009}; \citealt{Svoboda2015}).  This is supported by the spectral and timing analyses of \cite{Gallo2021}, where they also conclude that the data were consistent with obscuration from a single, homogeneous cloud. 

In this work, we will apply a colour-colour analysis to the data from the 2016 eclipse in NGC 6814.  Such grids will allow us to examine the eclipse with better time resolution than the ``state-resolved’’ spectral analysis of \cite{Gallo2021}.  The generation of light curves is presented in Section 2.  In Section 3, we describe the creation of the colour-colour grids.  In Section 4, we explore different geometries of the absorber(s) that might explain the colour-colour analysis.  Results and conclusion of this analysis are then discussed in Section 5 and 6, respectively.

\section{Data Processing and light curves}

We use archival data obtained from the \textit{XMM-Newton} (\citealt{Jansen+2001}) Science Archive (XSA) for a long observation of NGC 6814 taken on 8 April 2016. Here we focus on the data collected by the EPIC pn (\citealt{Struder+2001}) detector, which were presented by \cite{Gallo2021}, following their data reduction steps to extract calibrated event lists from the \textit{XMM-Newton} Observation Data Files (ODFs) using the \textit{XMM-Newton} Science Analysis System (SAS) version 20.0.0. Using the \textsc{epiclccorr} task, we produced light curves in four broad band passes: $0.3-10~\mathrm{keV}$, $0.3-1~\mathrm{keV}$, $1-4~\mathrm{keV}$, and $4-10~\mathrm{keV}$. Each light curve was rebinned into 1 ks time bins, and subsequently smoothed using a 5-point (i.e. 5 ks window) moving average to extract a cleaner representation of the underlying variable trend. In the forthcoming analysis, we ignore the final $\sim15~\mathrm{ks}$ of the observation due to significant background flaring. We note that we only present the results for the smoothed light curves here, though our results and interpretation do not change when using the unaltered data.

\begin{figure*}
    \centering
    \includegraphics[width=\linewidth]{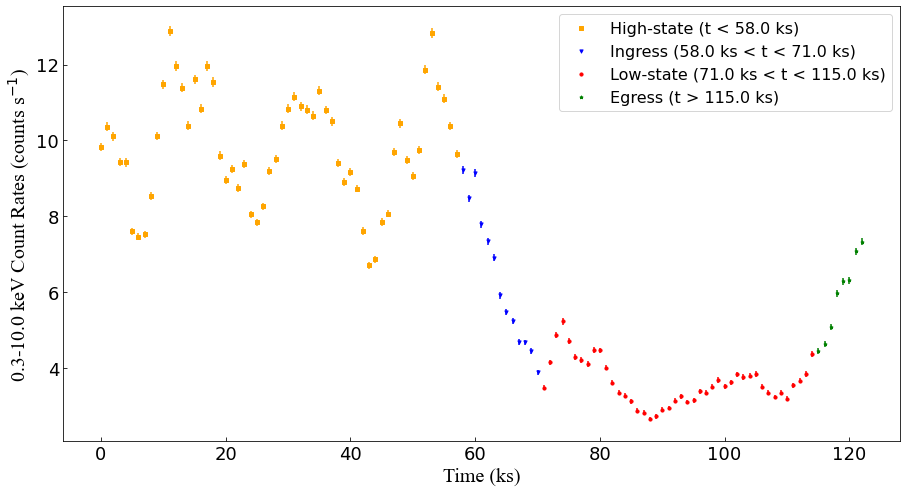}
    \includegraphics[width=\linewidth]{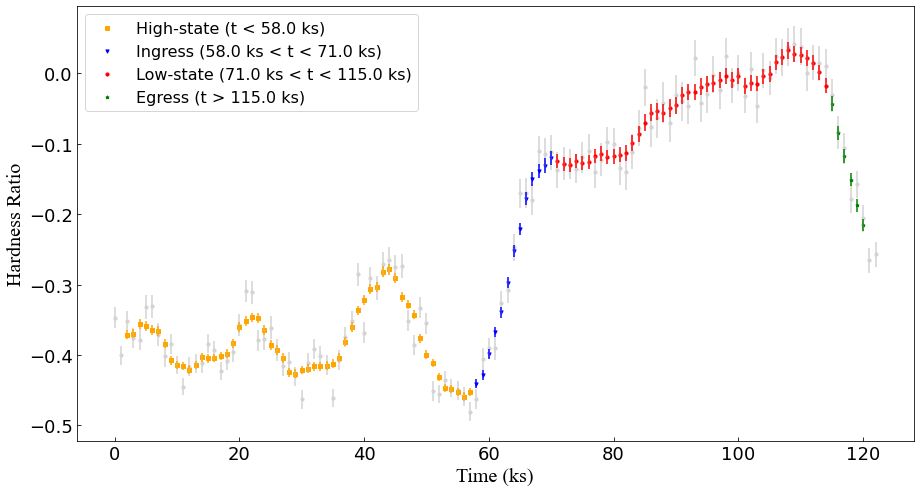}
    \caption{\textbf{Top Panel:} Light curve from the 2016 observation of NGC~6814 using the \textit{XMM-Newton EPIC-pn} detector, with counts in the $0.3-10.0~\mathrm{keV}$ energy band.  The uncertainties on the count rates are at the 1$\sigma$ statistical level. 
 \textbf{Bottom Panel:} Full HR curve (light gray) over-plotted with the smoothed HR, done by using moving average smoothing.  For HR, the bands used are S ($0.3-1.0~\mathrm{keV}$) and H ($4.0-10.0~\mathrm{keV}$), with $HR = (H - S)/(H + S)$.  The uncertainties on the HR are the 1$\sigma$ count rate errors propagated through the moving average smoothing.}
    \label{1}
\end{figure*}

\section{Colour-Colour Grids}

Colours can be used to study AGN variability by examining the effects of different model parameters (or parameter combinations) in colour-colour diagrams.  Comparing how a parameter is expected to change based on how the measured colour changes can reveal what is driving the variability (e.g. \citealt{Carpano2005}; \citealt{Grinberg2020}; \citealt{Nowak2011}).  In addition, colour changes can be measured with few photons so it is possible to examine variability and distinguish model degeneracies on much shorter time scales than with spectral modelling.

The count rates in each energy band depend on the telescope and instrument used, so colours are instrument dependent.   To determine the predicted colours, a model of interest is chosen and folded through the appropriate instrumental response files.  In this work, the {\it XMM-Newton}-pn responses generated in Section~2 are adopted.  

The model adopted here is based on the best-fit partial-covering model in \cite{Gallo2021} that describes the eclipsing event in NGC~6814.  The X-ray source was modeled as a power law ({\sc cutoffpl}) and blurred reflector ({\sc relxillD}), which was then modified by a variable partial coverer ({\sc zxipcf}).   All this emission was subject to two warm absorbers ({\sc xabsgrid}) (\citealt{Parker2019}; \citealt{Steenbrugge2003}) presumably originating at large distances.  In addition, a narrow Gaussian profile was included to model the narrow Fe~K$\alpha$ emission line at $\sim6.4~\mathrm{keV}$.  In {\sc xspec} jargon, the model appears as:  {\sc xabsgrid*xabsgrid * (zgauss + zxipcf (cutoffpl + relxillD))}.  \cite{Gallo2021} demonstrated that using a different continuum model did not significantly change the partial covering behaviour.

\begin{table}
	\centering
	\caption{The best-fit continuum model adopted from \citealt{Gallo2021} and the variable partial covering parameters used for creating the colour-colour grid (Fig.~\ref{2}).  In {\sc xspec} terminology, the best-fit model takes the form: {\sc xabsgrid*xabsgrid * (zgauss + zxipcf (cutoffpl + relxillD))}}
	\label{tab:example_table}
	\begin{tabular}{lccr} 
		\hline
		Model  & Model  & Parameter \\
		 Component &  Parameter &  Value\\
		\hline
		{\sc xabs$_{1}$} & log$\xi$/erg cm$^{-2}$ s$^{-1}$ & 2.81\\
		   & N$_{H}$/10$^{21}$ cm$^{-2}$ & 25.0\\
		   & v$_{out}$/km s$^{-1}$ & 76.0\\
		\hline
		{\sc xabs$_{2}$} & log$\xi$/erg cm$^{-2}$ s$^{-1}$ & 0.999\\
		   & N$_{H}$/10$^{21}$ cm$^{-2}$ & 3.59\\
		   & v$_{out}$/km s$^{-1}$ & 50.0\\  
            \hline
		{\sc zgauss} & E/keV & 6.45\\
		   & $\sigma$/eV & 137\\
		   & Norm/10$^{-5}$ ph. cm$^{-2}$ s$^{-1}$ & 5.41\\ 
            \hline
		{\sc zxipcf} & logN$_{H}$/cm$^{-2}$ & 18 values: [21, 24]\\
		   & log$\xi$/erg cm$^{-2}$ s$^{-1}$ & 1.09\\
		   & $f_{c}$ & 0.56, and 9 values: [0.0, 0.8]\\ 
            \hline
		{\sc cutoffpl} & $\Gamma$ & 1.99\\
            \hline
		{\sc relxillD} & q$_{in}$ & 8.48\\
		   & q$_{out}$ & 3\\ 
		   & R$_{b}$/$r_{g}$ & 6\\ 
		   & a/[cJ/GM$^{2}$] & 0.998\\ 
		   & i/$^{\circ}$ & 67\\ 
		   & log$\xi$/erg cm$^{-2}$ s$^{-1}$ & 0.326\\ 
		   & log N$_{H}$/ cm$^{-2}$ & 19\\ 
		   & A$_{Fe}$ & 3.39\\ 
            \hline
	\end{tabular}
\end{table}

In Fig.~\ref{1}, we see clear evidence of an eclipse.  The top panel shows the light curve in the $0.3-10.0~\mathrm{keV}$ energy band, with the transient behaviour starting around $60~\mathrm{ks}$, where the brightness appears to exhibit the ingress, minimum, and egress stages of an eclipse.   The HR curve alongside it (bottom panel) points to absorption as the source of the variability.  Concurrent with the decrease in count rate, there is a noticeable increase in HR ("hardening"), which is a signature of absorption.  

\begin{figure*}
    \centering
    \includegraphics[width=\linewidth]{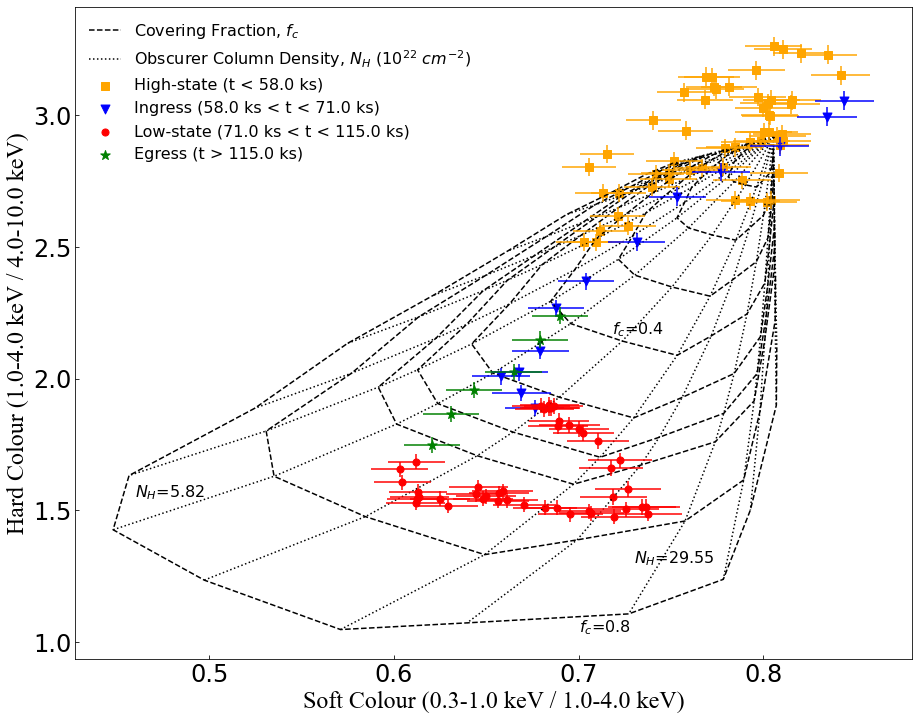}
    \includegraphics[width=\linewidth]{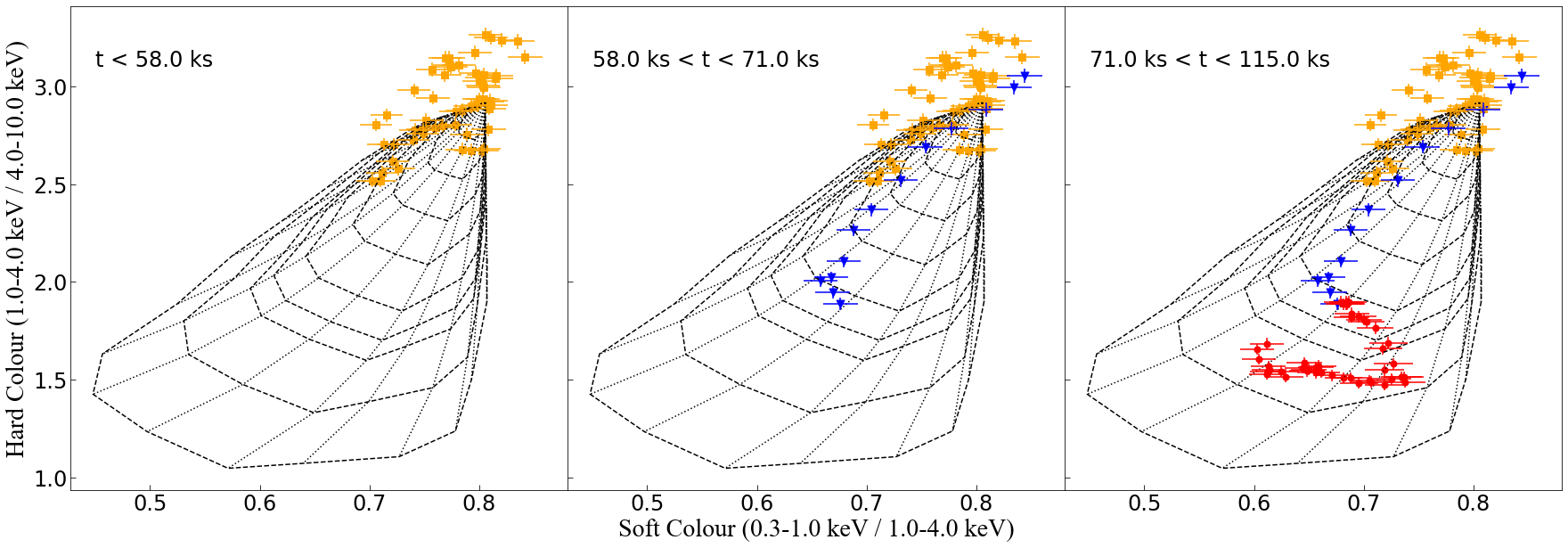}
    \caption{Colour-colour grid constructed from varying covering fraction and column density, plotted with data from NGC 6814.  Curves of constant column density and constant covering fraction are shown in dotted and dashed lines, respectively.  \textbf{Top Panel:} The complete figure with all the data plotted.  The data are shown sequentially in the lower panels. \textbf{Lower-left Panel:} The high-flux pre-flare data. \textbf{Lower-mid Panel:} High-flux and ingress. \textbf{Lower-right Panel:} High-flux through to the end of the eclipse maximum.  The egress is shown in the top panel.  An animation portraying the evolution of the colours through time is available in the online version of the journal. The uncertainties on the colours are the 1$\sigma$ count rate errors propagated through the moving average smoothing.}
    \label{2}
\end{figure*}

The important component in this work is the variable partial covering absorber, since changes in its covering fraction ($f_c$, fraction of the X-ray source that is covered by the cloud) and column density ($N_H$) were sufficient to replicate the eclipse in NGC~6814.  We vary these two parameters in our base model and extract the count rates in three specific energy bands to create the two colours used in the colour-colour diagram.  Specifically, the two colours are:

\begin{equation}
    C1 = S/M
\end{equation}
and
\begin{equation}
    C2 = M/H,
\end{equation}
where $S=0.3-1.0~\mathrm{keV}$,  $M=1.0-4.0~\mathrm{keV}$, and $H=4.0-10.0~\mathrm{keV}$.  By varying through different combinations of covering fraction and column density, we create a grid in colour-colour space depicted by curves of constant covering fraction (red curves) and constant column density (blue curves) (Fig.~\ref{2}).

In Fig.~\ref{2}, the data from NGC~6814 and the partial-covering grid are over-plotted in colour-colour space.  The colour combination shows a clear transition through four different states of the eclipse (e.g. high-state, ingress, low-state and egress).  An animation portraying the time progression of the colours is available in the online version of the journal.

The light curve and hardness ratio curve of NGC~6814 exhibit substantial fluctuations during the high-state prior to the eclipse ($t \leq 58$~ks) (Fig.~\ref{1}).  In Fig.~\ref{2}, there is significant scatter in colour-colour space during the high-state (black data).  This intrinsic variability is interesting and the subject of current work in preparation.  Regarding the eclipse analysis, the intrinsic variations are not incorporated in the grids.  Attempts to do so did not provide improvements over considering the obscuration alone.  The changes after 60 ks are dominated by the eclipse.  This could be because the high-flux variations appear to driven by changes in the soft band, which are completely obscured during the eclipse (in prep).

During ingress ($t \approx 58-71$ ks), the changes in Fig.~\ref{2} (cyan data points), are dominated by changes in covering fraction, with only a slight shift in column density.  With increasing time, the data points follow a curve of roughly constant column density ($N_H\sim 5.82$) while the covering fraction changes from 0 to $\sim 0.8$.  The pattern is almost exactly reversed during egress ($t \geq 115$ ks) as the data move along the same track in the opposite direction, i.e. decreasing covering fraction with approximately constant column density. 

The low state (red data, $t \approx 71-115$ ks) shows when the X-ray source is most highly eclipsed / obscured.  In the light curve (Fig.~\ref{1}), the count rate during this phase is roughly constant.  However, the hardness ratio curve (bottom panel, Fig.~\ref{1}) displays a steady hardening during the low state even though the count rate is not varying significantly.   

The colour-colour diagram (Fig.~\ref{2}) can be used to explain this behaviour.  While in ingress, the changes were dominated by the covering fraction increasing, during the first part of the low-state, the column density also begins to increase.  At approximately $90$~ks, the covering fraction and column density are both at their maximum values.  This also corresponds to the lowest flux in the light curve (Fig.~\ref{1}).  After this point, while still in the low-state of the eclipse, the covering fraction remains constant while the column density begins to decrease.

\begin{figure*}
    \centering
    \includegraphics[scale=0.385]{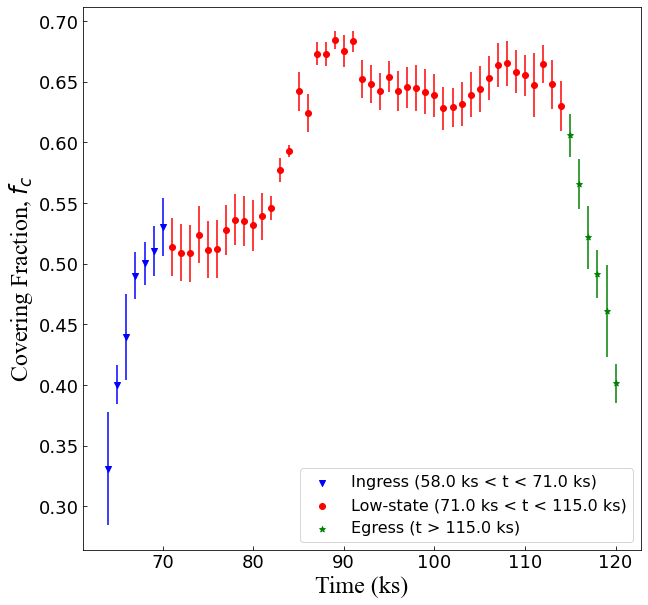}
    \includegraphics[scale=0.385]{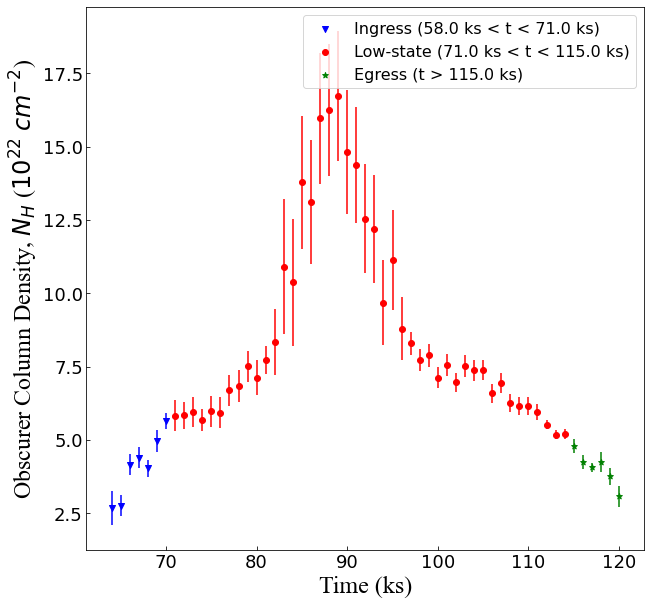}
    \caption{The changes in the covering fraction (\textbf{left}) and column density (\textbf{right}) as a function of time for the eclipse in NGC~6814.  The curves cover the eclipse period between $60-120$~ks and the colours match the stages defined in Fig.~\ref{1} and ~\ref{2}. }
    \label{3}
\end{figure*}
The covering fraction and column density changes with time are presented in Fig.~\ref{3}.
In the left panel, the covering fraction curve shows the expected steady increase and decrease during ingress and egress, respectively.  There is a brief flattening in covering fraction (i.e. covering fraction) evident at the start of the low state followed by a sharp rise at $\sim85$~ks to a peak value.  This is followed by a slight drop and further flattening until the end of the low-state (perhaps another peak at $\sim110$~ks).  

The right panel of Fig.~\ref{3} depicts the column density curve with time.  From the start of ingress, there is a continuous, though not consistent rise to a maximum column density during the low-state.  Similarly, from maximum column density there is a continuous, though not consistent drop to egress.  In addition, the peak in column density does not occur at the mid-point of the eclipse.  Understanding the variations in the covering fraction and column density of the obscurer during the low state is particularly important as it might reveal the nature of the medium.

To extract the column density and covering fraction for each data point, we employ the use of the {\sc shapely} package in {\sc python}.  Since most points on the grid are contained between two curves of constant covering fraction and two curves of constant column density, their positions on the grid can give approximate values for those two parameters.  We interpolate between these curves of constant column density and covering fraction to obtain an ($N_H, f_C$) pair of values for each point on the grid. To determine the uncertainties on the interpolated values, the same processes is followed taking into consideration the relative uncertainty on the colour.  Effectively, the $1\sigma$ uncertainties on the colours are translated to uncertainties on the ($N_H, f_C$) pair.

\begin{figure*}
    \centering
    \includegraphics[scale=0.385]{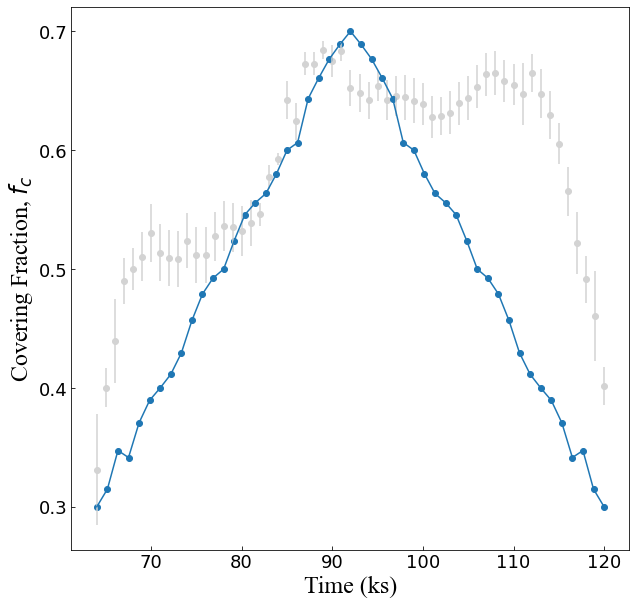}
    \includegraphics[scale=0.385]{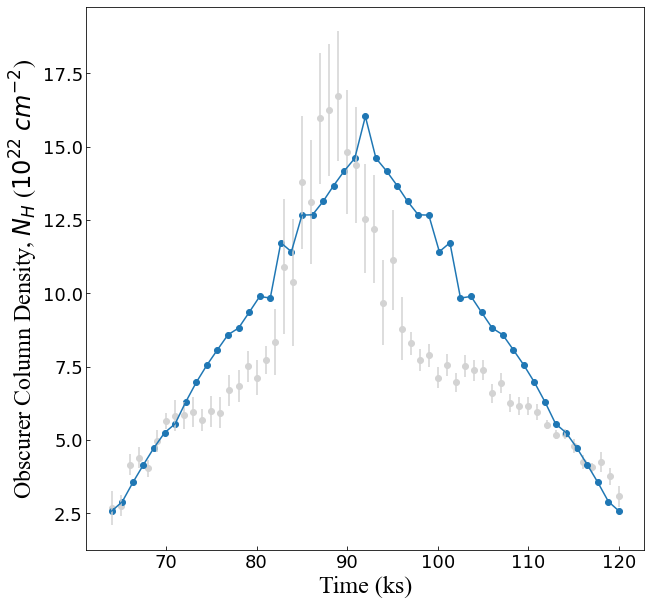}
    \includegraphics[width=\linewidth]{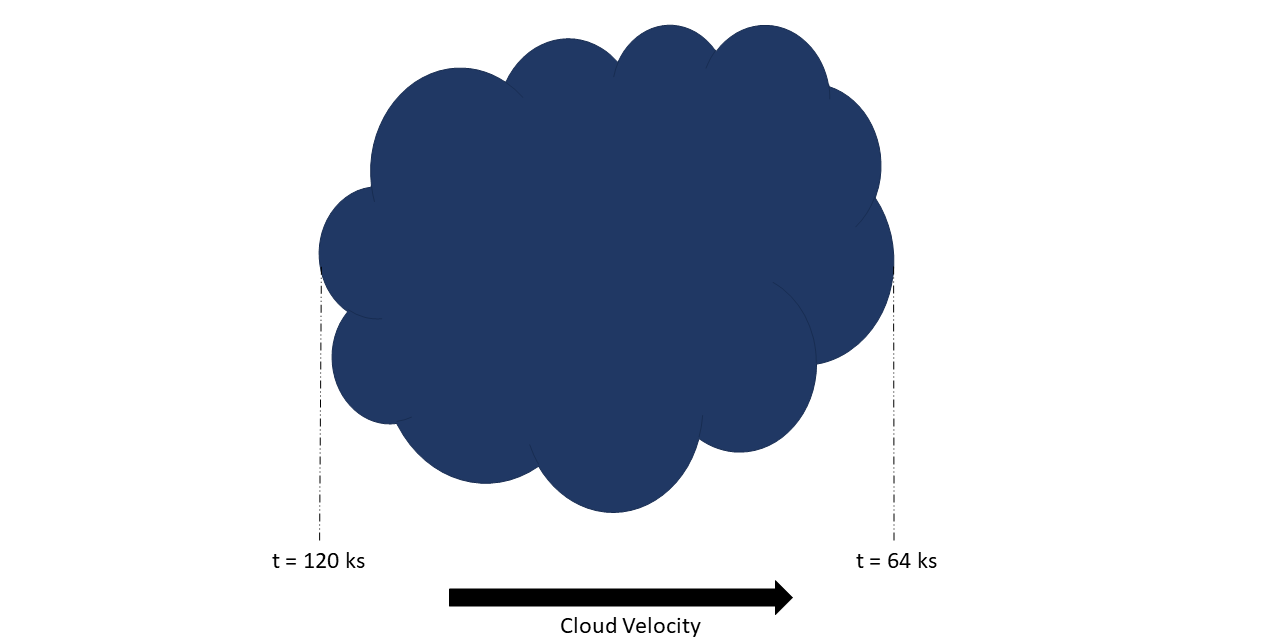}
    \caption{Simulation consisting of a single, homogeneous cloud showing the expected changes in covering fraction (\textbf{top left}) and column density (\textbf{top right}) as a function of time.  A rough sketch of the single cloud system is also presented (\textbf{bottom}), showing how it would appear while crossing our line-of-sight (direction denoted by the cloud velocity).  Dashed lines indicate start and end times.  The simulated cloud varies in $f_{c}$ from 0.3 to 0.7 back to 0.3, and varies in $N_{H}$ from 2.5 to 15.0 back to 2.5, in units of 10$^{22}$ $cm^{-2}$.}
    \label{4}
\end{figure*}
The complexity in the shapes of both curves shown in Fig.~\ref{3} indicate the eclipse was not from the passing of a single, homogeneous cloud (Fig.~\ref{4}).  Fig.~\ref{4} presents the covering fraction and column density change predicted from simulations of single, homogeneous cloud passing in front of the X-ray source.   For the simulation, the cloud is assumed to have a minimum column density and covering fraction at the beginning of the eclipse, reach a maximum in those parameters at the middle of the eclipse (i.e., the lowest flux/most absorbed state), and return to the minimum values at the end of the eclipse.  The figure shows a uniform triangular shape for both curves clearly consistent with a simple increase to decrease scenario as would happen with such a cloud.  This preliminary simulation, thus, shows that an alternate geometry is required to explain the eclipse, as will be discussed later in this paper.

\section{Using More Realistic Modelling of the Obscurer}

In addition to the single cloud obscurer, we considered three other scenarios to attempt to describe the behaviour in Fig.~\ref{3}.  Specifically, we looked at a double cloud model, a triple cloud model, and finally three clouds embedded in a highly ionized halo.

The objective is not to perfectly reproduce the covering fraction and column density changes seen in NGC~6814 (Fig.~\ref{3}), but rather to determine if such contrived obscurers could create some of the behaviour seen in Fig.~\ref{3}, like the flattening in the covering fraction curve or the asymmetric peak in the column density curve.

\subsection{Double cloud}

\begin{figure*}
    \centering
    \includegraphics[scale=0.385]{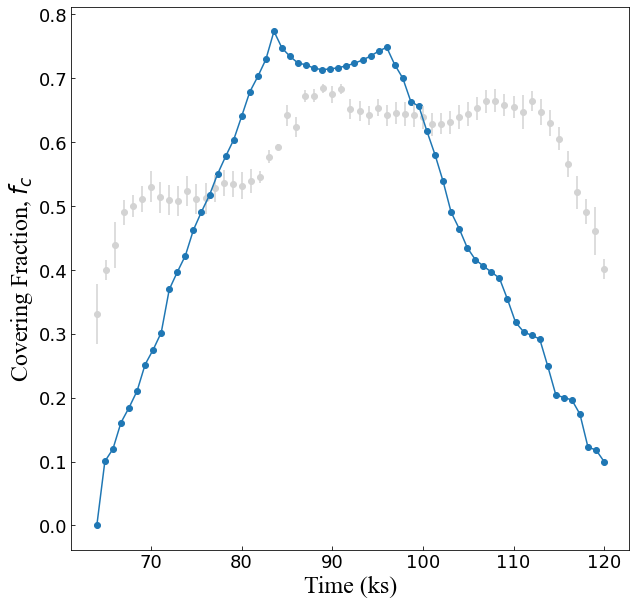}
    \includegraphics[scale=0.385]{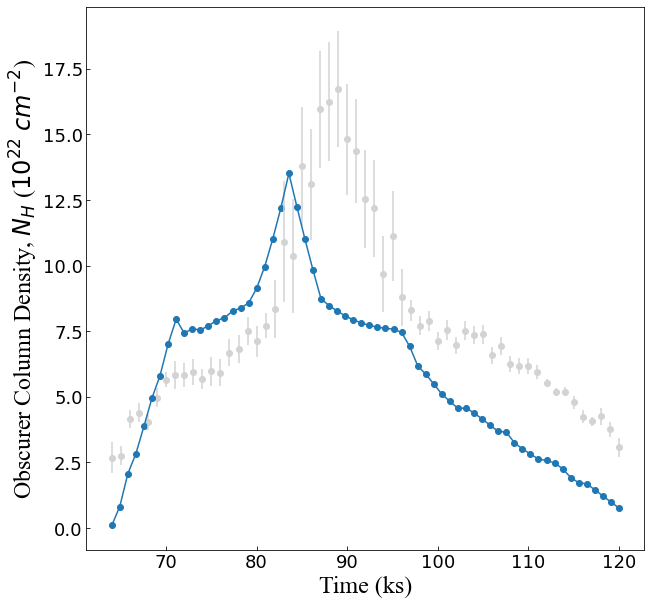}
    \includegraphics[width=\linewidth]{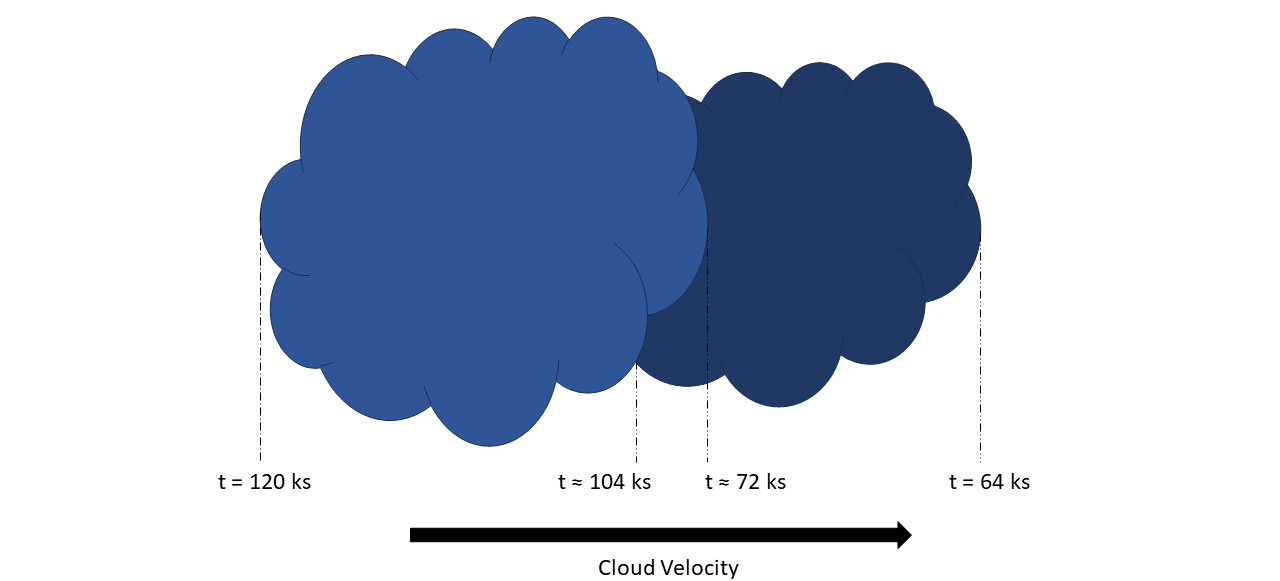}
    \caption{Simulation consisting of two homogeneous clouds showing the expected changes in covering fraction (\textbf{top left}) and column density (\textbf{top right}) as a function of time.  The placements of the clouds in the simulation are arbitrary and will affect the overall shape of the curves.  A rough sketch of the double cloud system is also presented (\textbf{bottom}), showing how it would appear while crossing our line-of-sight (direction denoted by the cloud velocity).  Dashed lines indicate start and end times for both clouds.  The darker shaded cloud represents a cloud with a higher maximum column density, and the relative sizes of the clouds relate to how their covering fractions compare.  The first cloud (64 ks $\le$ t $\le$ 104 ks) varies in $f_{c}$ from 0.1 to 0.7 back to 0.1, and varies in $N_{H}$ from 1.0 to 21.0 back to 1.0, in units of 10$^{22}$ $cm^{-2}$.  The second cloud (72 ks $\le$ t $\le$ 120 ks) varies in $f_{c}$ from 0.1 to 0.65 back to 0.1, and varies in $N_{H}$ from 0.8 to 6.0 back to 0.8, in units of 10$^{22}$ $cm^{-2}$.}
    \label{5}
\end{figure*}

In simulating the double cloud scenario, we consider a system consisting of two homogeneous clouds that partially overlap.  Each cloud is modeled with a {\sc zxipcf} component that has a variable covering fraction and column density.  The ionisation parameter ($\xi = L/nr^{2}$, with $L$ being the ionizing luminosity of the source, and $n$ being the cloud density a distance $r$ from the source) of each cloud is fixed at $log(\xi) = 1.08733$.  The degree to which the clouds overlap is arbitrary.  In the simulation shown in Fig.~\ref{5}, the clouds overlap for $30.625~\mathrm{ks}$.  

The inclusion of a second cloud results in two peaks in the covering fraction curve corresponding to the times when each cloud covers the most.  In between the peaks, the covering fraction curve flattens (Fig.~\ref{5}, left panel).  The multiple cloud scenario also introduces asymmetry in the column density curve (Fig.~\ref{5}, right panel) where the peak column density is reached when the path-length through both clouds is maximum.

The exact location of the peaks and the duration of the flattening is dependent on the placement of the two clouds in the simulation.  A larger separation would result in more distinct peaks.  However, the flattening and asymmetry does resemble the behaviour seen in the original eclipse data (Fig.~\ref{3}).  These simulations are not exact replications of the data, but appear to be a step in the right direction compared to the simulations of a single, homogeneous cloud.

The scenario described is not strictly representative of two homogeneous clouds, but could be applied more generally.  The simulations could also be replicating non-spherical and / or non-homogeneous obscurers.

\subsection{Triple cloud}

\begin{figure*}
    \centering
    \includegraphics[scale=0.385]{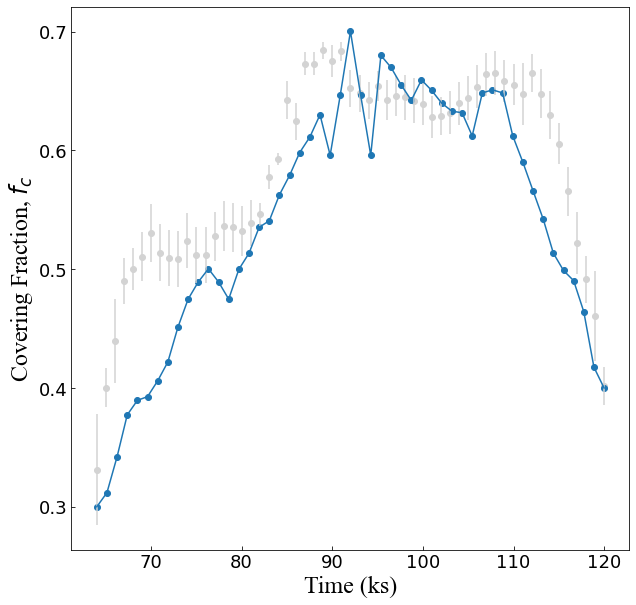}
    \includegraphics[scale=0.385]{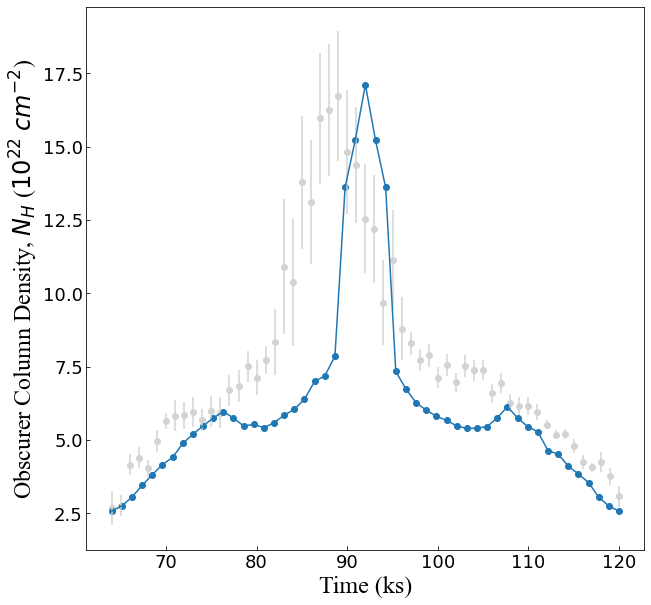}
    \includegraphics[width=\linewidth]{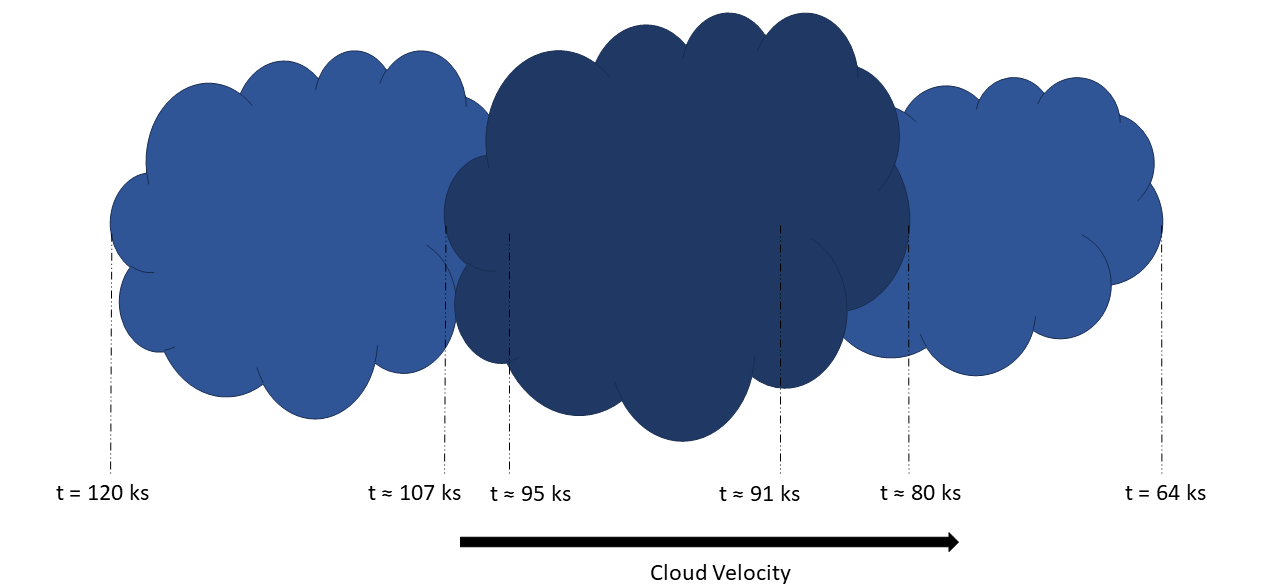}
    \caption{Simulation consisting of three homogeneous clouds showing the expected changes in covering fraction (\textbf{left}) and column density (\textbf{right}) as a function of time.  The placements of the clouds in the simulation are arbitrary and will affect the overall shape of the curves.  A rough sketch of the triple cloud system is also presented (\textbf{bottom}), showing how it would appear while crossing our line-of-sight (direction denoted by the cloud velocity).  Dashed lines indicate start and end times for each of the clouds.  The darker shaded cloud represents a cloud with a higher maximum column density than the other two, and the relative sizes of the clouds relate to how their covering fractions compare.  The first cloud (64 ks $\le$ t $\le$ 91 ks) varies in $f_{c}$ from 0.3 to 0.5 back to 0.3, and varies in $N_{H}$ from 2.5 to 6.0 back to 2.5, in units of 10$^{22}$ $cm^{-2}$.  The second cloud (80 ks $\le$ t $\le$ 107 ks) varies in $f_{c}$ from 0.1 to 0.7 back to 0.1, and varies in $N_{H}$ from 5.0 to 16.0 back to 5.0, in units of 10$^{22}$ $cm^{-2}$.  The third cloud (95 ks $\le$ t $\le$ 120 ks) varies in $f_{c}$ from 0.4 to 0.66 back to 0.4, and varies in $N_{H}$ from 2.5 to 6.0 back to 2.5, in units of 10$^{22}$ $cm^{-2}$.}
    \label{6}
\end{figure*}

A three homogeneous cloud system that arbitrarily overlap each other is a simplification of a multi-cloud obscurer.  In Fig.~\ref{6}, we present the simulation of such a system and find it replicates the behaviour in NGC~6814 well.  The addition of this third obscurer results in significant fluctuations in the parameters over time.  Present in the covering fraction curve are multiple regions of flattening and two peaks.  Additionally, sharp rises and drops are also evident, similar to the data.  The column density curve nearly shows multiple peak, a large peak with small peaks on either side.  This indicates that a multi-cloud geometry could be plausible for describing the obscurer in explaining in NGC~6814.

\subsection{Multiple clouds in an ionised coma}

\begin{figure*}
    \centering
    \includegraphics[scale=0.385]{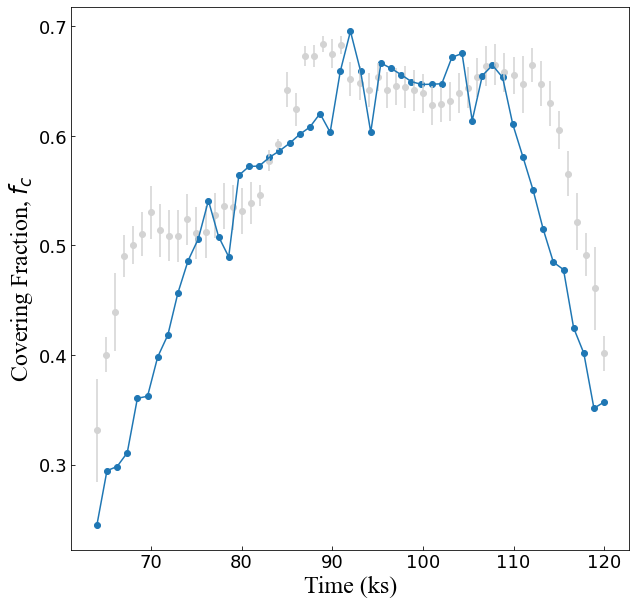}
    \includegraphics[scale=0.385]{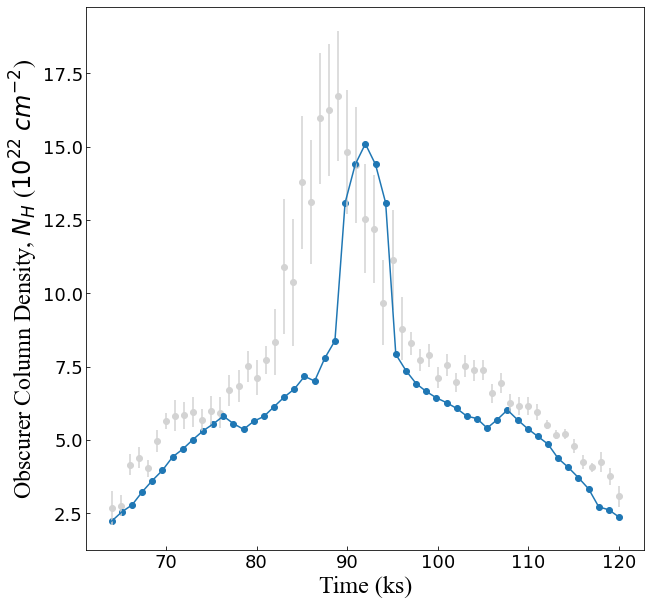}
    \includegraphics[width=\linewidth]{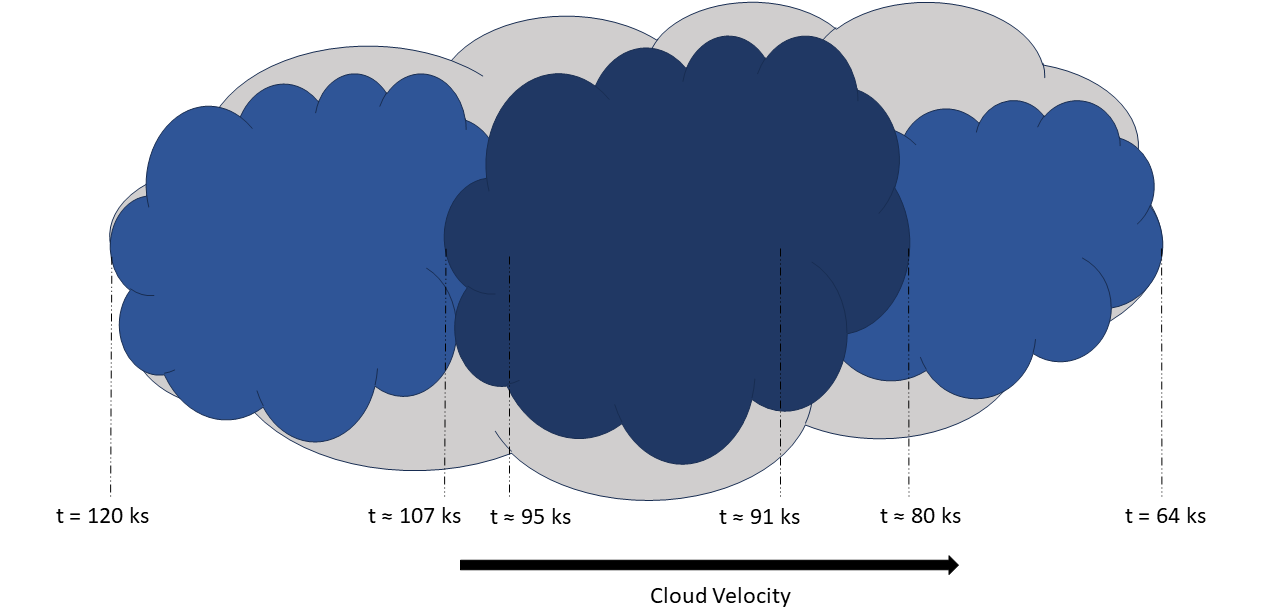}
    \caption{Simulation consisting of three homogeneous clouds, plus some higher ionization cloud/halo showing the expected changes in covering fraction (\textbf{left}) and column density (\textbf{right}) as a function of time.  The placements of the clouds in the simulation are arbitrary and will affect the overall shape of the curves.  A rough sketch of the triple cloud + higher-ionization cloud system is also presented (\textbf{bottom}), showing how it would appear while crossing our line-of-sight (direction denoted by the cloud velocity).  Dashed lines indicate start and end times for each of the clouds.  The darker shaded cloud represents a cloud with a higher maximum column density than the others, and the relative sizes of the clouds relate to how their covering fractions compare.  The higher-ionization ($log(\xi) = 2.0$) cloud (spanning the entire time) is shown in a lighter colour than the others to distinguish it.  The first cloud (64 ks $\le$ t $\le$ 91 ks) varies in $f_{c}$ from 0.2 to 0.5 back to 0.2, and varies in $N_{H}$ from 2.5 to 6.0 back to 2.5, in units of 10$^{22}$ $cm^{-2}$.  The second cloud (80 ks $\le$ t $\le$ 107 ks) varies in $f_{c}$ from 0.2 to 0.67 back to 0.2, and varies in $N_{H}$ from 6.0 to 17.0 back to 6.0, in units of 10$^{22}$ $cm^{-2}$.  The third cloud (95 ks $\le$ t $\le$ 120 ks) varies in $f_{c}$ from 0.3 to 0.65 back to 0.3, and varies in $N_{H}$ from 2.5 to 6.0 back to 2.5, in units of 10$^{22}$ $cm^{-2}$.  Additionally, the higher-ionization cloud spans the entire period, and varies in $N_{H}$ from 0.1 to 0.6 back to 0.1, in units of 10$^{22}$ $cm^{-2}$, while staying constant at $f_{c}$=0.6.}
    \label{7}
\end{figure*}

The final simulation is the three-cloud system (as above), but the clouds are now embedded in a large coma or halo of highly ionised material.  The ionisation level of the coma is fixed at $log(\xi)=2.0$, which is comparable to the degree of ionisation found in \cite{Gallo2021}.  The simulation is shown in Fig.~\ref{7}.

The differences between this simulation and the three-cloud system without the ionised coma are subtle.  The distinct features in the covering fraction curve are still present, but the ingress ($t\sim 60-70$~ks) and egress ($t\sim 110-120$~ks) stages are less steep as the values are slightly larger.  The main difference is in the column density curve, where the ingress and egress stages are more rounded and better replicate the observation.  

The simulations demonstrate that the eclipsing obscurer in NGC~6814 was more complex than a single, homogenous cloud.  Some compact, clumpy cloud system, perhaps embedded within a highly ionised coma, could describe the eclipsing in NGC~6814.

\section{Discussion}

\begin{figure*}
    \centering
    \includegraphics[width=\linewidth]{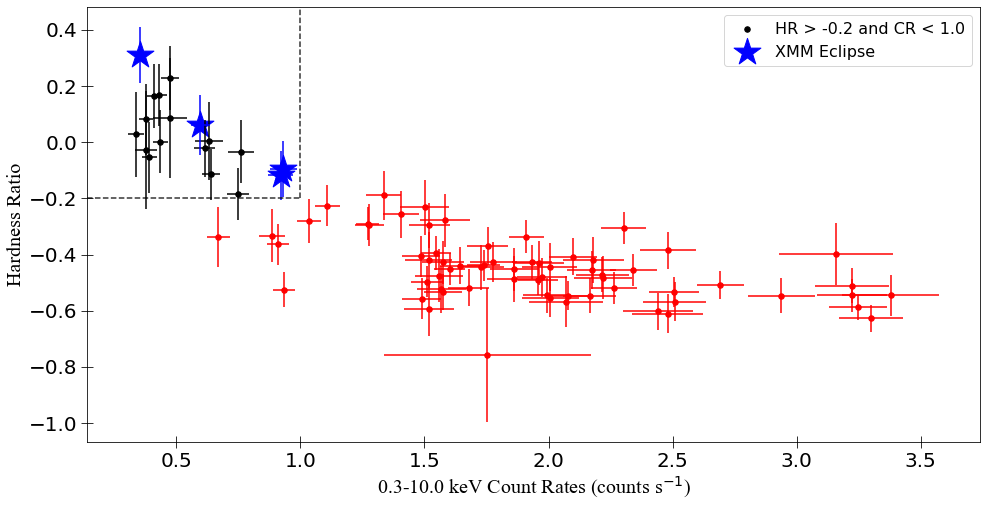}
    \includegraphics[width=\linewidth]{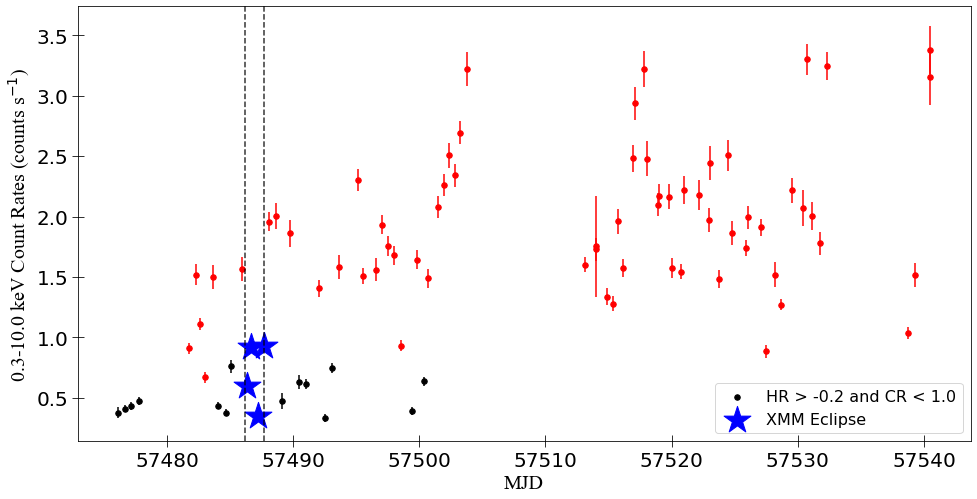}
    \caption{\textit{Swift} XRT monitoring of NGC~6814 for $\approx 64$~days around the time of the \textit{XMM-Newton} eclipse. The hardness ratio (HR) as a function of count rate (CR) is shown in the top panel where data points with HR > $-0.2$ and CR < $1.0$ are sectioned off and shown in black.  The blue stars correspond to observations that occurred during the \textit{XMM-Newton} eclipse.  The points in the box meet the eclipse criteria (see text for details).  For HR, the bands used are S ($0.3-1.0~\mathrm{keV}$) and H ($4.0-10.0~\mathrm{keV}$), with $HR = (H - S)/(H + S)$.
    The \textit{Swift} XRT light curve ($0.3-10.0~\mathrm{keV}$) is plotted in the lower panel. The period during the \textit{XMM-Newton} eclipse is marked by vertical dashed-lines.  Using the count rate errors from the product builder, Gaussian error analysis is used to calculate the hardness ratio errors.}
    \label{8}
\end{figure*}

Eclipsing events have been caught in several AGN (e.g. \citealt{DeMarco2020}; \citealt{Constanzo2022}; \citealt{Brenneman2013}; \citealt{Turner2018}; \citealt{Nardini2011}; \citealt{Kaastra2014}; \citealt{Longinotti2013}; \citealt{Longinotti2019}; \citealt{Ebrero2016}; \citealt{Risaliti2007}; \citealt{Risaliti2011}).  Even though eclipses are transient, X-ray observations of them provide diagnostics of the absorbing medium and X-ray source, which is important as such regions are impossible to image with current detectors.  

In the case of NGC~6814, studying the eclipse in colour-colour space reveals complexity in the nature of the obscurer.  The variation in the column density and covering fraction, both of which connect to its geometry of the obscurer passing the line-of-sight, indicate something more complicated than a simple, single cloud.  We gain important information on the location and nature of the obscuring medium, which in turn helps us better understand the AGN environment and the importance of AGN winds. 

The event in NGC~6814 is unique amongst eclipsing AGN, as the 2016 \textit{XMM-Newton} observation captured the entire eclipse, which lasted about one day.  \textit{Swift} also captured the eclipse, as it observed  NGC~6814 for $\sim60$-days over a similar time frame. The \textit{Swift} light curve, showing the long-term variability over two-months is in the bottom panel of Fig.~\ref{8}.  The \textit{Swift}  observations occurring during the \textit{XMM-Newton} pointing are identified as stars in the figure.  The data were downloaded from the \textit{Swift} XRT Product Builder (\citealt{Evans2009})\footnote{\url{https://www.swift.ac.uk/user_objects/}}.  

Additionally, the top panel of Fig.~\ref{8} portrays the \textit{Swift} hardness ratio as a function of count rate.  Again, the data coincident with the \textit{XMM-Newton}  pointing are identified with blue stars.  Using the \textit{XMM-Newton} eclipse as a baseline, we can establish an eclipse criteria. The verified \textit{XMM-Newton}  eclipse is clearly distinguished with the criteria of HR > -0.2 and CR < 1.0 (sectioned off by the dashed lines).  In addition, high-cadence observations of NGC~6814 in 2022 indicate that HR$ < -0.2$ are consistent with the unobscured state of the AGN (Gonzalez et al. in prep).

Based on the established hardness ratio – count rate criteria, there are several points (black points in Fig.~\ref{8}) in the $\sim2$ month \textit{Swift} light curve that resemble the behaviour during the eclipse suggesting there may have been several obscuring events in 2016.   All these events occur in the $\sim30$~days prior to MJD = 57505. Based on previous reports about NGC~6814 (e.g. \citealt{Gallo2021}; \citealt{Leighly1994}), eclipses in this source may not be uncommon, which is further supported by the possibility that these fluctuations indicate it being prone to more eclipses during the 2016 epoch.  If the black points in Fig.~\ref{8} are associated with eclipses, this would imply a large covering factor of the sky as seen from the black hole.

With this eclipse criteria in mind, we can get an idea of the number of obscuring clouds in the line-of-sight during the \textit{Swift} observation.  Finding the occasions when the data suggest the onset and end of an eclipse (i.e. shifting from red to black to red data points), we identify about 6.5 possible eclipsing events.  Since the first event occurs right at the beginning of the light curve, and it only shows an increase, it is possible that this is not a full eclipsing event, thus, we consider it half of an eclipse.  Using this information, and the fact that various properties of the partial coverer were calculated in \cite{Gallo2021}, we can estimate the fraction for the occurrence of these events, both throughout the Swift observation and the period of heavy eclipsing activity (before the gap at MJD = 57505~days of Fig.~\ref{8}).

From \cite{Gallo2021}, the Keplerian velocity of the partial coverer was calculated to be $V_{K} = 10^{4} km/s$, the location of it was calculated to be $r = 2694~r_{g} = 4.34 \times 10^{15}$~cm.  This results in an orbital pathlength of $C = 2\pi r = 2.73 \times 10^{16}$~cm, and an orbital period (Keplerian timescale) of $T_{K} = C/V_{K} = 27269$~ks.  The duration of the entire \textit{Swift} light curve is $t_{lc} \sim 64~\rm days  = 5529.6$~ks, which covers approximately $0.2$ of the Keplerian orbit.

The estimated diameter of the cloud was calculated to be $D_{C} = 1.30 \times 10^{13}$~cm. Assuming all the cloud sizes are similar and that there are 6.5 eclipsing events, the total distance (arc length) covered by the clouds would be $D_{tot} = 6.5~D_{C} = 8.45 \times 10^{13}$~cm.  Based on this, the fraction of the orbit covered by clouds is $\approx 0.015$. 

All the events occur prior to MJD = 57505 days.  The duration of the \textit{Swift} light curve during this phase is $\sim 27~\rm days = 2332.8$~ks, corresponding to approximately $0.086$ of the Keplerian orbit.  The orbital covering factor during this heavy-obscuration phase is $\approx 0.036$. The \textit{Swift} light curve suggests that eclipses in NGC~6814 are likely and that the distribution of clouds in the orbit is not isotropic. 

In the original paper, \cite{Gallo2021} modelled the spectra during the high, ingress, low, and egress states.  The uniform light curve and time-resolved spectroscopy were indicative of a single cloud transiting the X-ray source.  As the analysis was mainly spectroscopic, time resolution was sacrificed for signal-to-noise. With the colour-colour analysis here, we can examine the transit with much better time resolution.  With the ability to examine the obscurer during the deepest part of the eclipse, we find that the obscurer may be inconsistent with a single homogeneous cloud.  As seen in Fig.~\ref{4}, the changes in both Cf and Nh are not symmetric, which is what would be expected for a uniform obscurer such as a single cloud.  Moving in this direction by testing various scenarios, it appears that the transit can be described by multiple clouds moving together.  This scenario (as depicted in both Fig.~\ref{6} and Fig.~\ref{7}) reproduces the Cf and Nh behaviour.    

The colour-colour analysis can be used to reveal details about the obscuring medium.  For NGC~6814, the obscuring is either inhomogeneous or substantially fragmented into many pieces.  A recent reverberation mapping study from a 2022 X-ray-to-optical monitoring campaign (Gonzalez et al. submitted) suggests that the outer disc is truncated at a smaller radius than is normally expected in such AGN.  This indicates a potential source for the obscuring material originating much closer to the black hole and might imply a higher rate of eclipses in NGC~6814 and the presence of inhomogenous material.

\section{Conclusion}

In this paper, we have used colours to examine the eclipsing event in NGC~6814.  We have found that these data exhibit absorption variability consistent with changes in both obscurer covering fraction and column density.  These variations are inconsistent with what would be expected from a single, homogeneous cloud.  Rather, more complex simulations showed that it is likely due to multiple clouds moving together.  

Long-term, high cadence monitoring of NGC~6814 would be useful to determine how common such eclipses are in the AGN.  This would reveal the nature and origin of the obscurer(s) causing the eclipses.  Additionally, this work could lead to better identifying eclipses in AGN, and characterising the behaviour of the obscurers.  

\section*{Acknowledgements}

This work was based on observations obtained with XMM-Newton, an ESA science mission with instruments and contributions directly funded by ESA Member States and NASA.  We also used data that was supplied by the UK Swift Science Data Centre (UKSSDC).  We would like to thank the referee for providing a constructive report that helped improve the paper. LCG acknowledge financial support from the Natural Sciences and Engineering Research Council of Canada (NSERC) and from the Canadian Space Agency (CSA).

\section*{Data Availability}

All data products used here are publicly available from the XMM-Newton Science Archive (XSA) and the UK Swift Science Data Centre (UKSSDC). 

\bibliographystyle{mnras}
\bibliography{papers}




\bsp	
\label{lastpage}
\end{document}